\documentclass[a4paper, 10pt, conference]{ieeeconf}      

\IEEEoverridecommandlockouts

\usepackage{cite}
\usepackage{amsmath,amssymb,amsfonts}
\usepackage{algorithmic}
\usepackage{graphicx}
\usepackage{textcomp}
\usepackage{xcolor}
\usepackage[normalem]{ulem} 
\usepackage{etoolbox, environ}
\usepackage{cleveref}
\usepackage[caption=false,font=footnotesize]{subfig}
\usepackage{fixltx2e}
\usepackage{siunitx}
\usepackage{xcolor}
\usepackage{todonotes} 
\usepackage{floatrow}
\Crefformat{figure}{#2Fig.~#1#3}
\Crefmultiformat{figure}{Figs.~#2#1#3}{ and~#2#1#3}{, #2#1#3}{ and~#2#1#3}

\newtoggle{is_draft}

\togglefalse{is_draft}

\NewEnviron{draft}
  {\iftoggle{is_draft}{\BODY}{}}

\begin{document}

\title{PoisSolver: a Tool for Modelling Silicon Dangling Bond Clocking Networks\\
\thanks{This work was supported by a research grant from the Natural Sciences and Engineering Research Council of Canada (NSERC), with funding reference number STPGP 478838-15.}
}

\author{
    Hsi Nien Chiu,
    Samuel S. H. Ng,
    Jacob Retallick,
    Konrad Walus, \textit{Member, IEEE}
}


\maketitle
\thispagestyle{plain}
\pagestyle{plain}

\begin{abstract}
Advancements in the fabrication of silicon dangling bonds (SiDBs) reveal a potential platform for clocked field coupled nanocomputing structures.
This work introduces PoisSolver, a finite element simulator for investigating clocked SiDB systems in the SiQAD design tool.
Three clocking schemes borrowed from prior work on quantum-dot cellular automata are examined as potential building blocks for a general clocking framework for SiDB circuits.
These clocking schemes are implemented in SiQAD, and power estimates are performed with geometrically agnostic methods to characterise each clocking scheme.
Clocking schemes using a \SI{14}{\nano\meter} technology node are found to dissipate \SI{10}{}-\SI{100}{\micro\watt\per\centi\meter\squared} at \SI{1}{\giga\hertz} and \SI{1}{}-\SI{10}{\watt\per\centi\meter\squared} at \SI{1}{\tera\hertz}.

\end{abstract}


\begin{draft}
  Example of using the draft markers.
\end{draft}

\section{Introduction} \label{sec_introduction}
Recent advancements in silicon dangling bond (SiDB) fabrication \cite{wolkow2014silicon, huff2017atomic, anderson2014field} presents a platform for beyond-CMOS binary logic.
A binary OR gate with dimensions less than $10\times\SI{10}{nm^2}$ \cite{huff2018binary} has been experimentally demonstrated on the SiDB platform, with similar simulation results shown for other common logic gates \cite{ng2020siqad}.
More complex SiDB circuitry requires data to be synchronized and directed through a large network of SiDBs.
To achieve this, the concept of clocking with suspended electrodes is borrowed from quantum-dot cellular automata (QCA) \cite{hennessy2001clocking, lent2003molecular}.
This work presents PoisSolver, a finite element simulator plugin to the SiQAD SiDB design tool \cite{ng2020siqad} for investigating clocked SiDB systems.

The paper is structured as follows.
\Cref{sec_background} provides an overview of the SiDB platform and prospective clocking techniques.
\Cref{sec_clock_schemes} introduces three selected clocking schemes serving as case studies.
\Cref{sec_poissolver} demonstrates PoisSolver's functionality in obtaining surface potential solutions for a clocked SiDB system.
\Cref{sec_circuit} models the clocking network as an RC circuit, and describes geometry agnostic methods for estimating resistance and capacitance.
\Cref{sec_power} analyses the power dissipation of the clocking electrode network.
Lastly, \Cref{sec_comparison} compares clocking power with other dissipative sources in the system.

\section{Background} \label{sec_background}
SiDBs are a novel approach to implement devices beyond the limits of CMOS fabrication, presenting potential advantages in device density, power, and speed.
SiDBs are created by selectively removing hydrogen atoms from hydrogen-passivated silicon surface dimers \cite{huff2017atomic}, and exhibit discrete charge states similar to quantum dots \cite{huff2018binary, taucer2014single}.
The platform relies on specific arrangements of SiDBs, encoding information positionally through charge state configurations \cite{wolkow2014silicon}.
OR gate experiments \cite{huff2018binary} demonstrate ground state tracking binary logic on the SiDB platform, with information manipulated through the placement and charge state of nearby SiDBs.
SiDB devices such as the OR gate can be switched without charge transfer, requiring only a reconfiguration of electrons and potentially lowering computational energy cost.
Simulation results of a full suite of binary logic gates are shown in \cite{ng2020siqad}, where the SiQAD design tool is used for the design and simulation of each gate.

The behaviour of SiDB gates relies on establishing specific electron populations and making desirable electron configurations energetically favorable under different input conditions.
SiDB placement and occupation is tweaked until ground state configurations mimic the truth table of a logical operation.
While past experiments demonstrate the use of scanning probes \cite{rashidi2016time} and surface contacts \cite{pitters2011charge} to control charge states, \cite{ng2020siqad} proposes that the transition levels (and thus the occupations) of the SiDBs can also be controlled by the application of an external field.
This work proposes the adaptation of suspended electrode clocking schemes found in QCA \cite{blair2018clock, vankamamidi2008two, campos2016use} for use in SiDB systems to implement control fields.

Clocked SiDB devices dissipate power through resistive losses in nonideal clocking networks in addition to any losses incurred within the SiDB system.
Quantitative analysis of clocking networks in the literature is limited to two phase networks using a simple series resistive capacitive model \cite{blair2010power}, neglecting parasitic self-capacitances and not addressing the four phase clocking network that many designs rely on.
To investigate clocking schemes in the context of SiDB systems, PoisSolver is designed for modelling electrode-silicon surface interactions specifically with the SiQAD design tool in mind.
PoisSolver is preferred over commercial tools such as COMSOL Multiphysics and ANSYS due to its integration with other physical simulators in the SiQAD environment.
We show the combined use of PoisSolver and SiQAD to extend power estimates of clocking networks using a four phase circuit model.

\section{Clock schemes under consideration} \label{sec_clock_schemes}
Clocking methods in QCA designs range from arbitrary and physically unrealistic \cite{liu2012review} to regular and fabricable \cite{blair2010power, vankamamidi2008two, campos2016use}.
This work will focus on the latter, targetting tileable clocking schemes for universal logic implementation.
Tileable schemes are selected for two major purposes: they are infinitely scalable for any SiDB netlist, and they predefine design constraints based on physically realisable clocking networks.

\begin{figure}
  \centering
  \begingroup 
    \captionsetup[subfigure]{width=0.5\linewidth}
    \subfloat[Columnar clocking pattern]{%
      \includegraphics[width=0.8\linewidth]{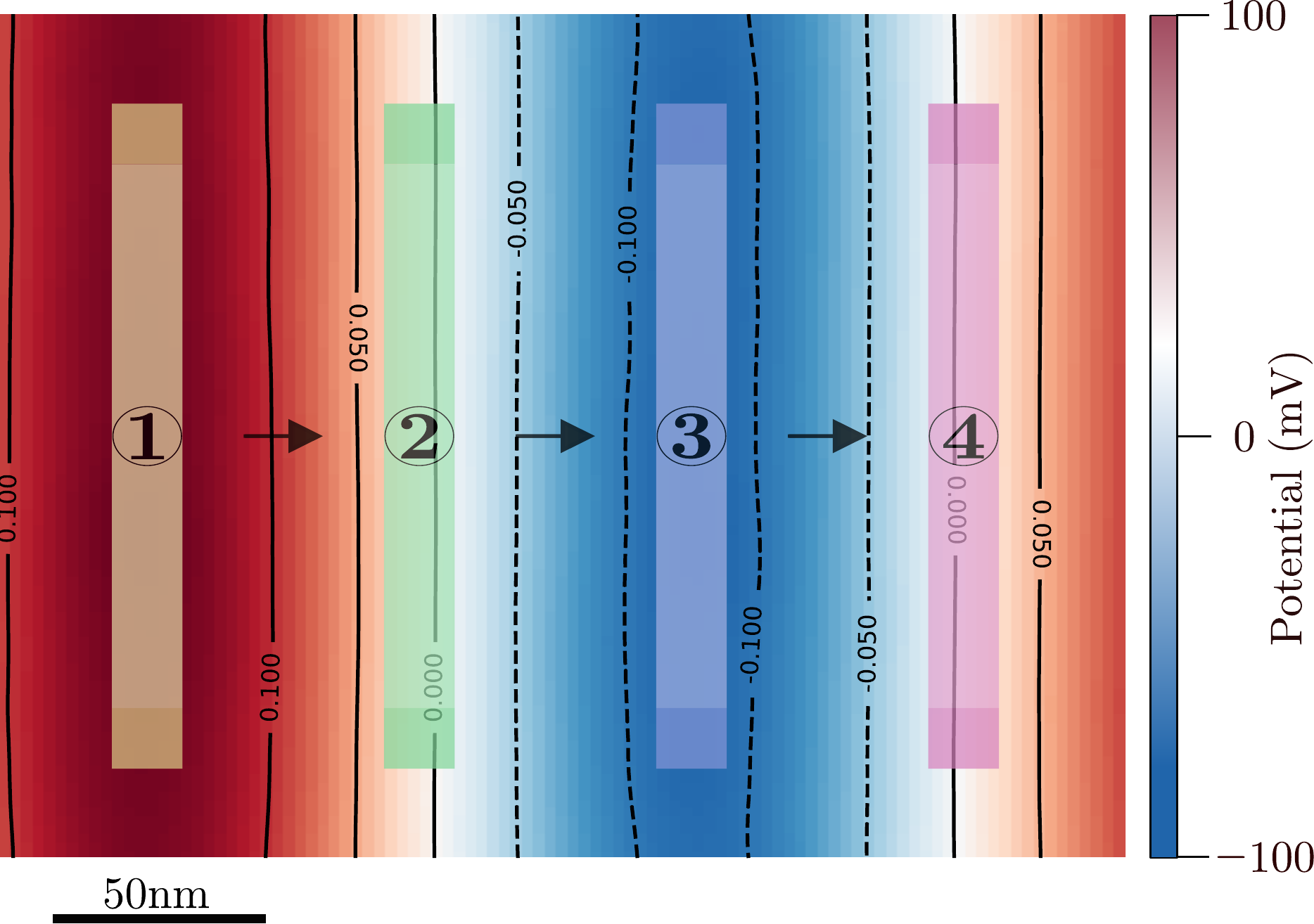}
      \label{sfig_col_clock}
    }
    \\
    \subfloat[Wave clocking pattern]{%
      \includegraphics[width=0.8\linewidth]{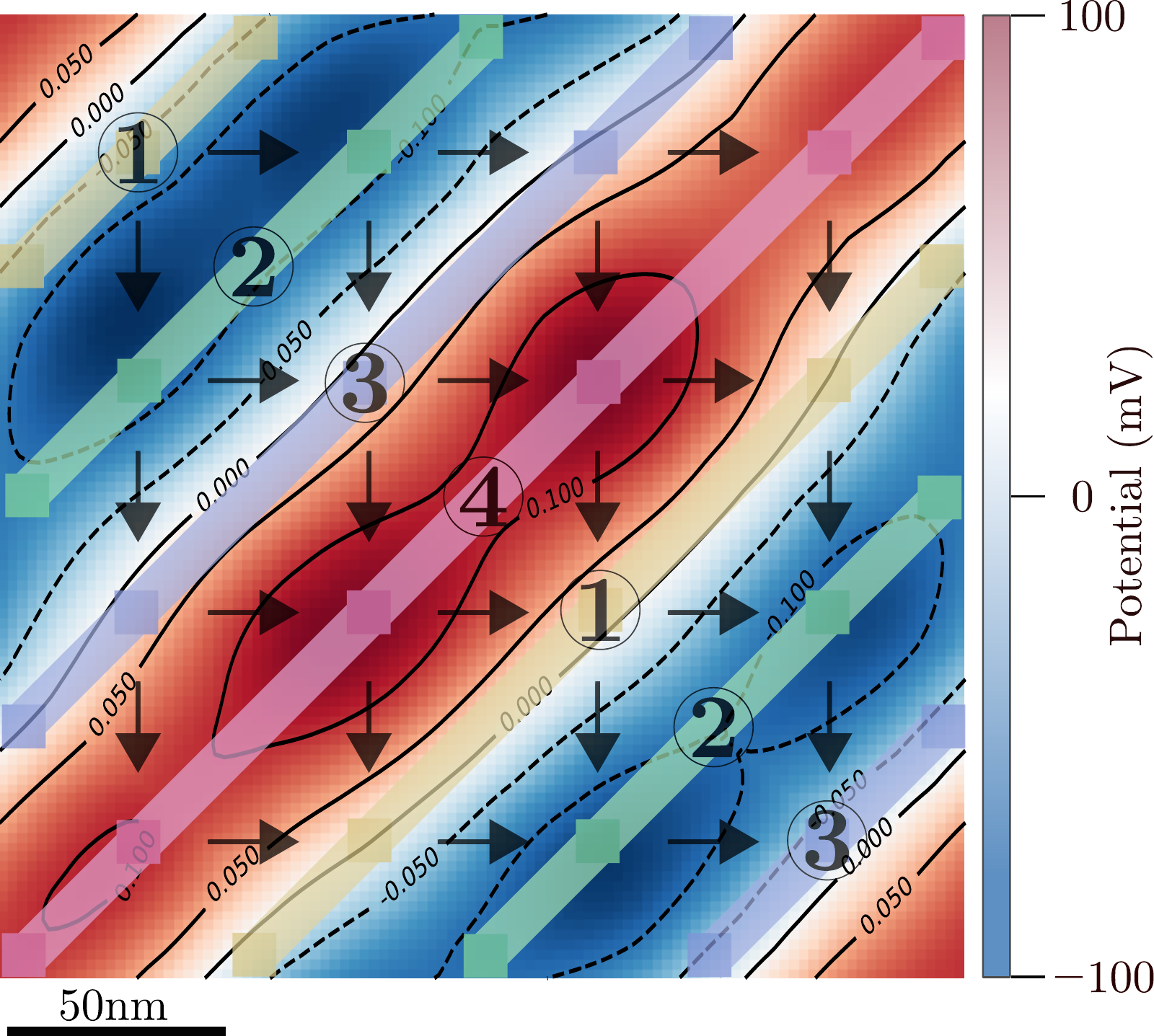}
      \label{sfig_wave_clock}
    }
    \\
    \subfloat[USE clocking pattern]{%
      \includegraphics[width=0.8\linewidth]{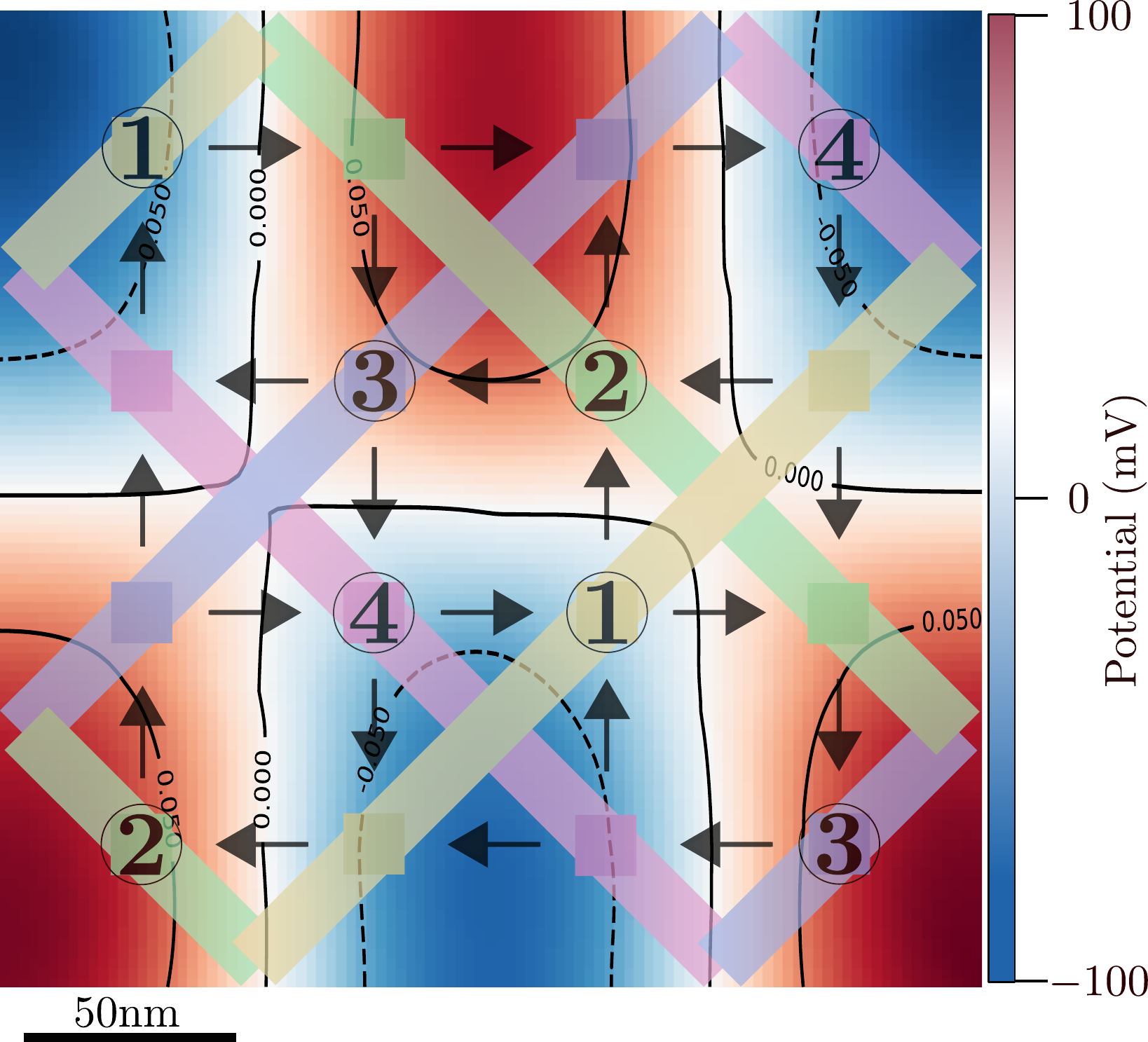}
      \label{sfig_use_clock}
    }
  \endgroup 
  \caption{Bird's eye view of single-tile electrode layouts for the columnar \protect\subref{sfig_col_clock}, wave \protect\subref{sfig_wave_clock}, and USE \protect\subref{sfig_use_clock} clocking schemes.
  Numbers denote electrode phases, and arrows indicate the direction of information flow.
  Colour map shows the potential swing on the silicon surface obtained by PoisSolver.
  Simulations assume periodic boundary conditions.
  }
  \label{fig:schemes_gui_layout}
\end{figure}

Three schemes are selected from literature: columnar \cite{blair2018clock}, wave \cite{vankamamidi2008two}, and USE \cite{campos2016use}.
SiQAD implementations of these schemes are shown in \Cref{fig:schemes_gui_layout}, along with numeric labels and arrows depicting relative phase and information flow respectively.
A \SI{14}{\nano\meter} technology node \cite{sicard2017introducing} was selected, yielding design parameters shown in \Cref{table_parameters}.
Following these geometric constraints, a single tile is created for each clocking scheme as compactly as possible.

\begin{table}
  \centering
  \begin{tabular}{ |c|c|c|  }
   \hline
   Parameter & Symbol & Value \\
   \hline
   Minimum metal pitch & $\lambda$ & \SI{53.7}{\nano\metre}\\
   Minimum metal width & w & \SI{14}{\nano\metre}\\
   Minimum metal thickness & t & \SI{10}{\nano\metre}\\
   \hline
  \end{tabular}
  \caption{Table of parameters for the system under consideration.}
  \label{table_parameters}
\end{table}

\section{PoisSolver} \label{sec_poissolver}
\begin{figure}
  \centering
  \includegraphics[width=0.85\linewidth]{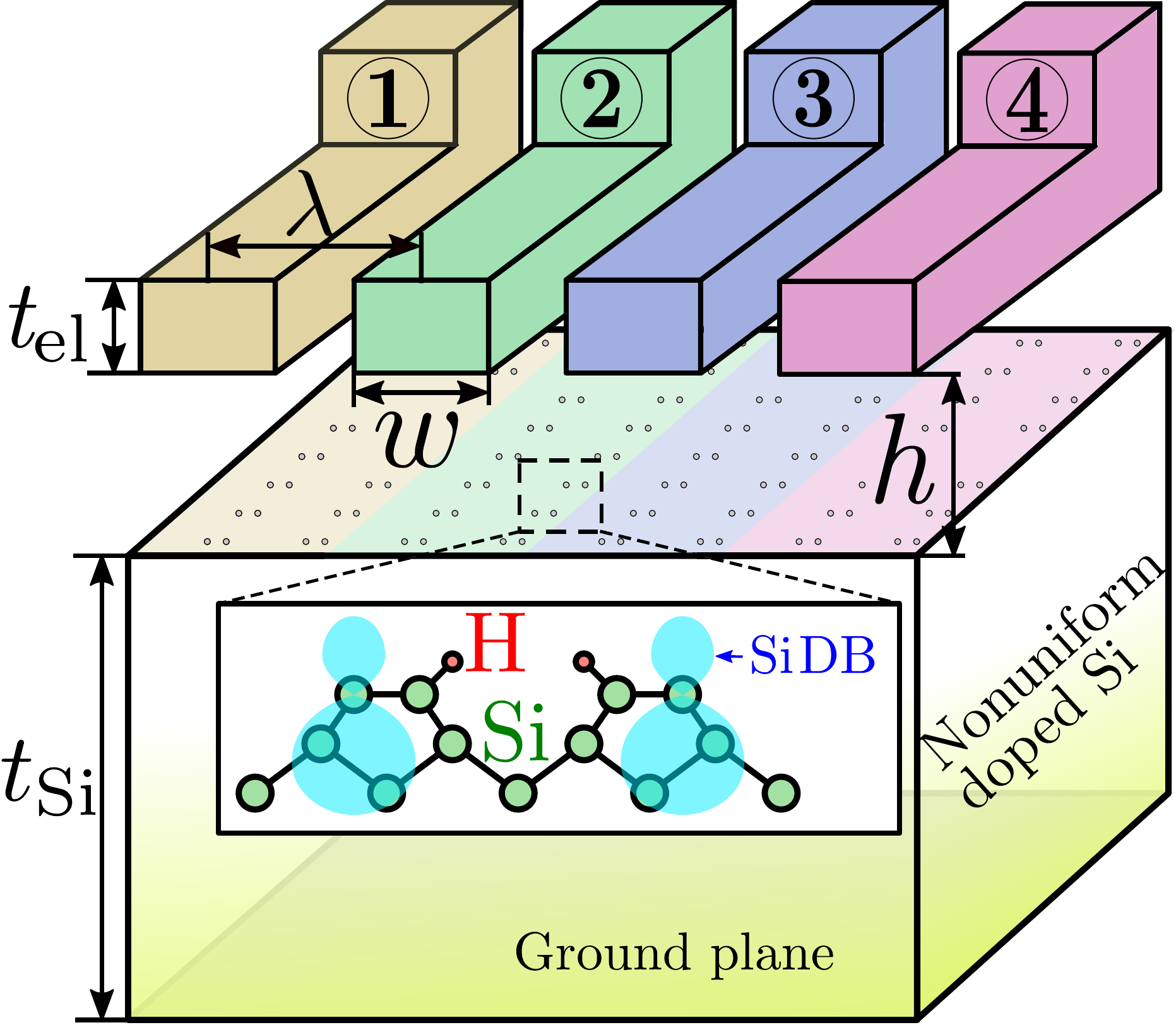}
  \caption{The schematic of the system under consideration (not to scale).
           The electrodes are implemented using CMOS metal layers with pitch ($\lambda$), width ($w$), electrode height ($h$), electrode thickness ($t_{\text{el}}$), and substrate thickness ($t_{\text{Si}}$).
           Electrode phases and clock zones are indicated by both numerical labels and colour.
           A close up schematic of SiDBs on the surface is also shown, depicting the H-passivated dimers and the SiDB orbital.
  }
  \label{fig:system_model}
\end{figure}

With the electrode network defined and routed in the SiQAD design environment, we consider the combined system of electrodes and SiDBs.
A model of the clocked SiDB system is shown in \Cref{fig:system_model}, consisting of a silicon substrate, SiDBs on a hydrogen-passivated surface, and the network of electrodes suspended in dielectric above the SiDBs.
SiDB clocking is achieved by modulating electrode voltages to induce desired surface potential patterns on the silicon surface to bias the flow of information.

\subsection{Finite element model} \label{subsec_physical_model}
To quantify these surface potential patterns in this 3D system we utilize a finite element model.
PoisSolver is developed in Python, and takes advantage of the FEniCS \cite{logg2010dolfin} \cite{alnaes2015fenics} finite element package.
Its primary function is to take user-defined electrode layouts in SiQAD and solve Poisson's Equation.
The electrode network is meshed using Gmsh \cite{geuzaine2009gmsh} via the Open CASCADE geometry kernel.
Formally, PoisSolver solves the generalized Poisson's Equation:
\begin{equation}
   -\nabla \cdot (\epsilon \nabla u) = \rho (x) \quad x \in \Omega \label{subeq_poisson}
\end{equation}
with the dielectric constant $\epsilon$, the three dimensional potential $u$, the spatial charge density $\rho$, the position $x$, and the simulation domain $\Omega$.
By altering $\rho$, Eq. \ref{subeq_poisson} can be adapted to account for different physical models.
PoisSolver supports four different definitions of $\rho$:
\begin{subequations}
  \begin{align}
    &\text{LE}: &\rho &= 0 \label{rho_laplace} \\
    &\text{PE}: &\rho &= q(N_D - N_A + p - n) \label{rho_poisson} \\
    &\text{PBE}: &\rho &= q(N_D - N_A + n_ie^{\frac{-qu}{kT}}- n_ie^{\frac{qu}{kT}}) \label{rho_pois_boltz} \\
    &\text{LPBE}: &\rho &= q(N_D + \frac{2q^2n_i}{kT}u) \label{rho_lin_pois_boltz}
  \end{align}
\end{subequations}
Laplace's Equation (LE) is used in the case of an intrinsic silicon substrate.
Poisson's Equation (PE) uses static definitions for $p$ and $n$, the hole and electron concentrations, accounting for a doped silicon substrate.
The Poisson-Boltzmann Equation (PBE) extends PE by including charge migration effects.
The Linearized Poisson-Boltzmann Equation (LPBE) is valid for systems with small $u$.
The PBE is the most physically representative of the system, but its nonlinearity results in long convergence times.

Appropriate boundary conditions are necessary to arrive at the expected solution.
Four types of boundary conditions are supported in PoisSolver:
\begin{subequations}
 \begin{align}
   u = f_D \quad x \in \Gamma_D \label{subeq_dirichlet} \\
   \epsilon n \cdot u = g_N \quad x \in \Gamma_N \label{subeq_neumann} \\
   \epsilon n \cdot u + h_R u = 0 \quad x \in \Gamma_R \label{subeq_robin} \\
   u(x+x_p) = u(x) \quad x_p \in \mathbb{R}^3 \label{subeq_periodic}
 \end{align}
\end{subequations}
where $\Gamma_D$, $\Gamma_N$, $\Gamma_R$ are the boundaries for the Dirichlet, Neumann, and Robin boundary conditions respectively, with boundary parameters $f_D$, $g_N$, and $h_R$.
Electrodes are defined as domains with high $\epsilon$ and Dirichlet boundary conditions on their surfaces.
All other boundary conditions are used on simulation boundaries for uniqueness.
\Cref{subeq_periodic} is the periodic boundary condition with periodicity vector $x_p$, and is used to mimic the infinite tiling of schemes discussed in \Cref{sec_clock_schemes}.

\subsection{Comparison between models} \label{subsec_compare_models}
\begin{figure}
  \centering
  \begingroup 
    \captionsetup[subfigure]{width=0.5\linewidth}
    \subfloat[LE]{%
      \includegraphics[width=0.425\linewidth]{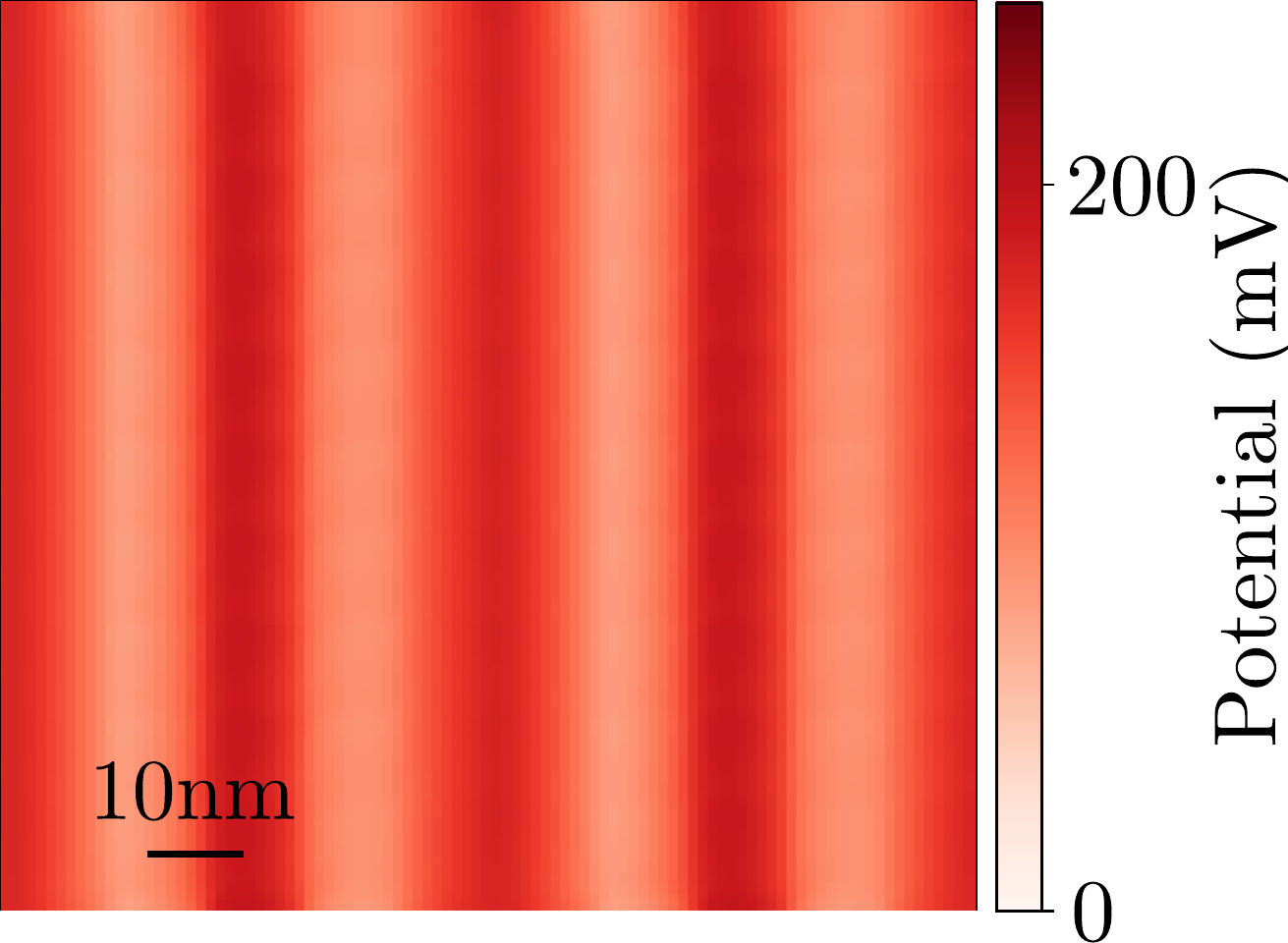}
      \label{sfig_le}
    }
    \quad
    \subfloat[PE]{%
      \includegraphics[width=0.425\linewidth]{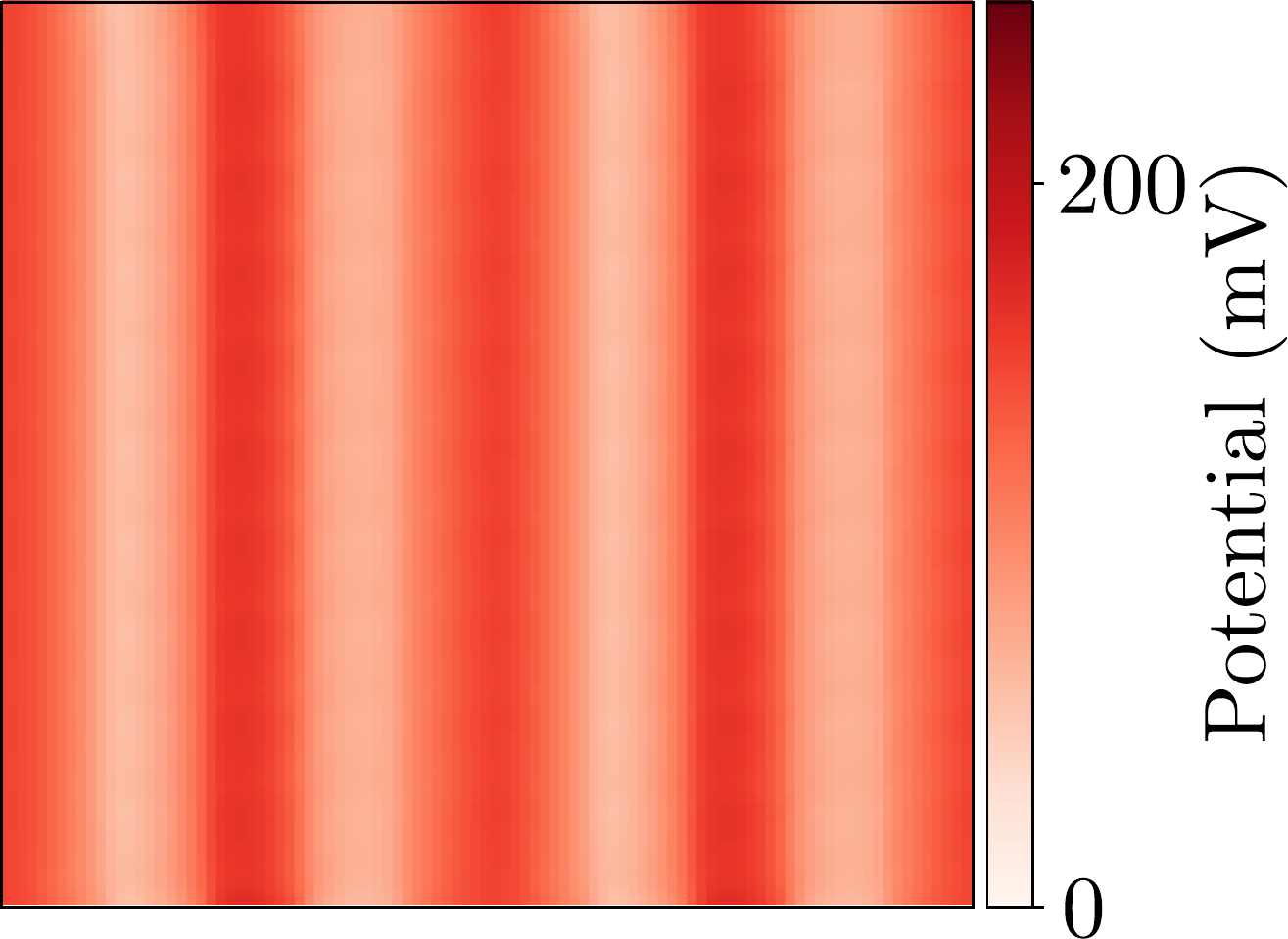}
      \label{sfig_pe}
    }
    \\
    \subfloat[PBE]{%
      \includegraphics[width=0.425\linewidth]{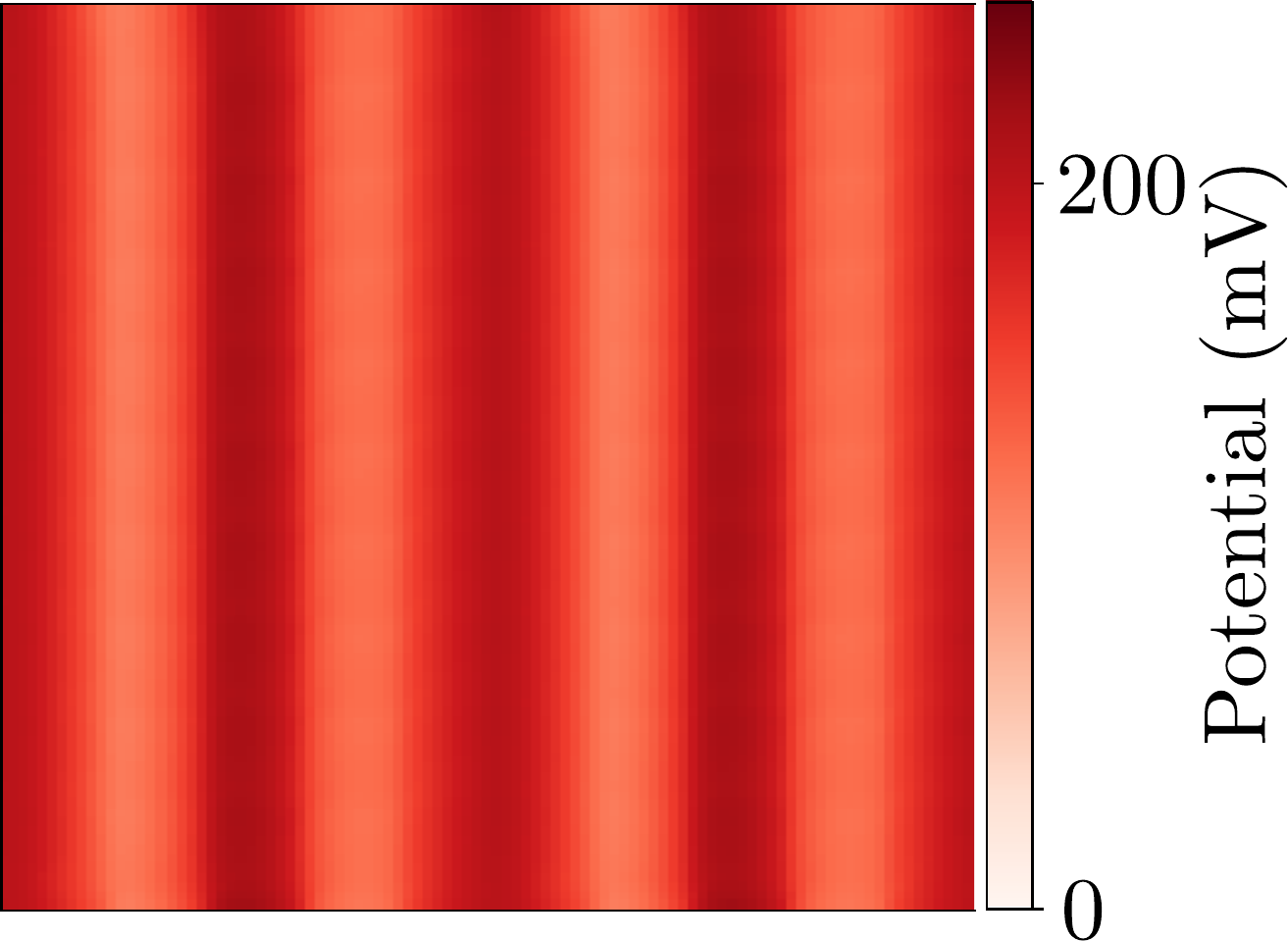}
      \label{sfig_pbe}
    }
    \quad
    \subfloat[LPBE]{%
      \includegraphics[width=0.425\linewidth]{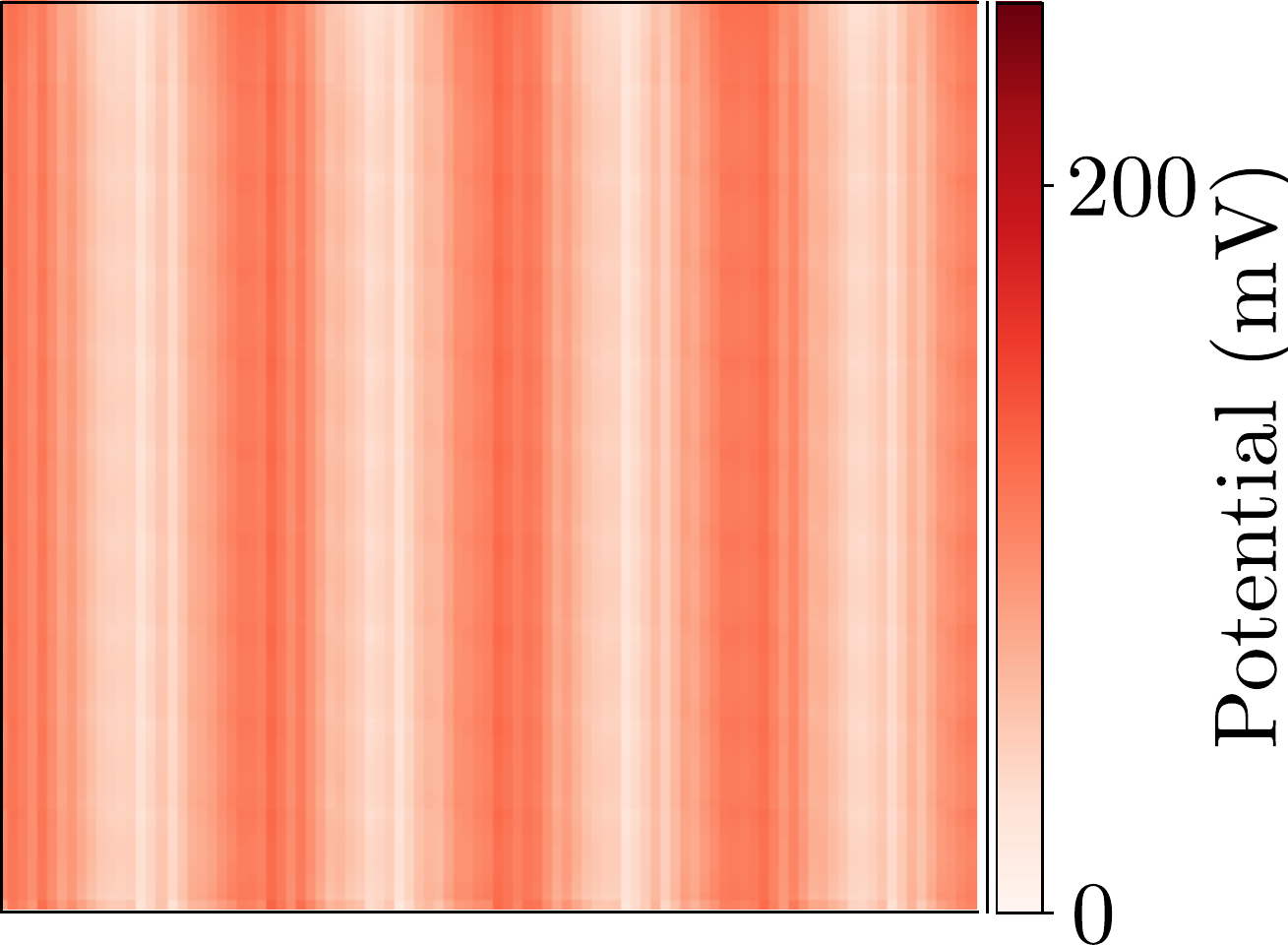}
      \label{sfig_lpbe}
    }
  \endgroup 
  \caption{Surface potential results from a simplified columnar system using different physical models.
           Although each physical model offers a slightly different potential solution, all supported physical models converge to solutions that resemble columnar clocking patterns.
  }
  \label{fig:model_slices}
\end{figure}

\Cref{fig:model_slices} shows potentials at the silicon surface for the columnar system using each of the four physical models with periodic boundary conditions in $x$ and $y$.
Each model converges to a columnar clocking pattern, with slightly different offsets and potential swings.
These offsets are attributed to the way each model handles spatial charge density.

\begin{figure}
  \centering
  \begingroup 
    \captionsetup[subfigure]{width=0.5\linewidth}
    \subfloat[ALE vs. PBE]{%
      \includegraphics[width=0.45\linewidth]{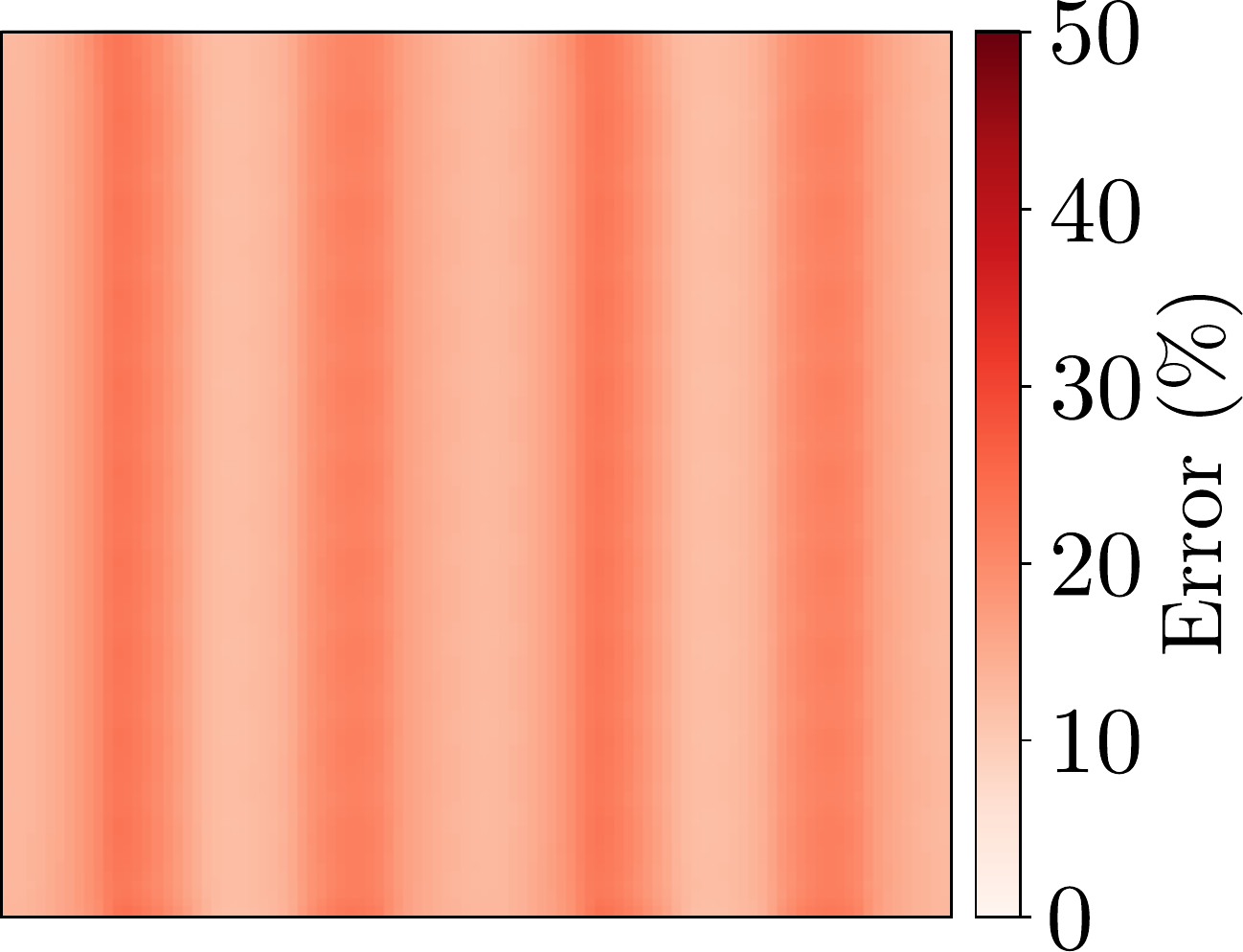}
      \label{sfig_le_error}
    }
    \quad
    \subfloat[PE vs. PBE]{%
      \includegraphics[width=0.45\linewidth]{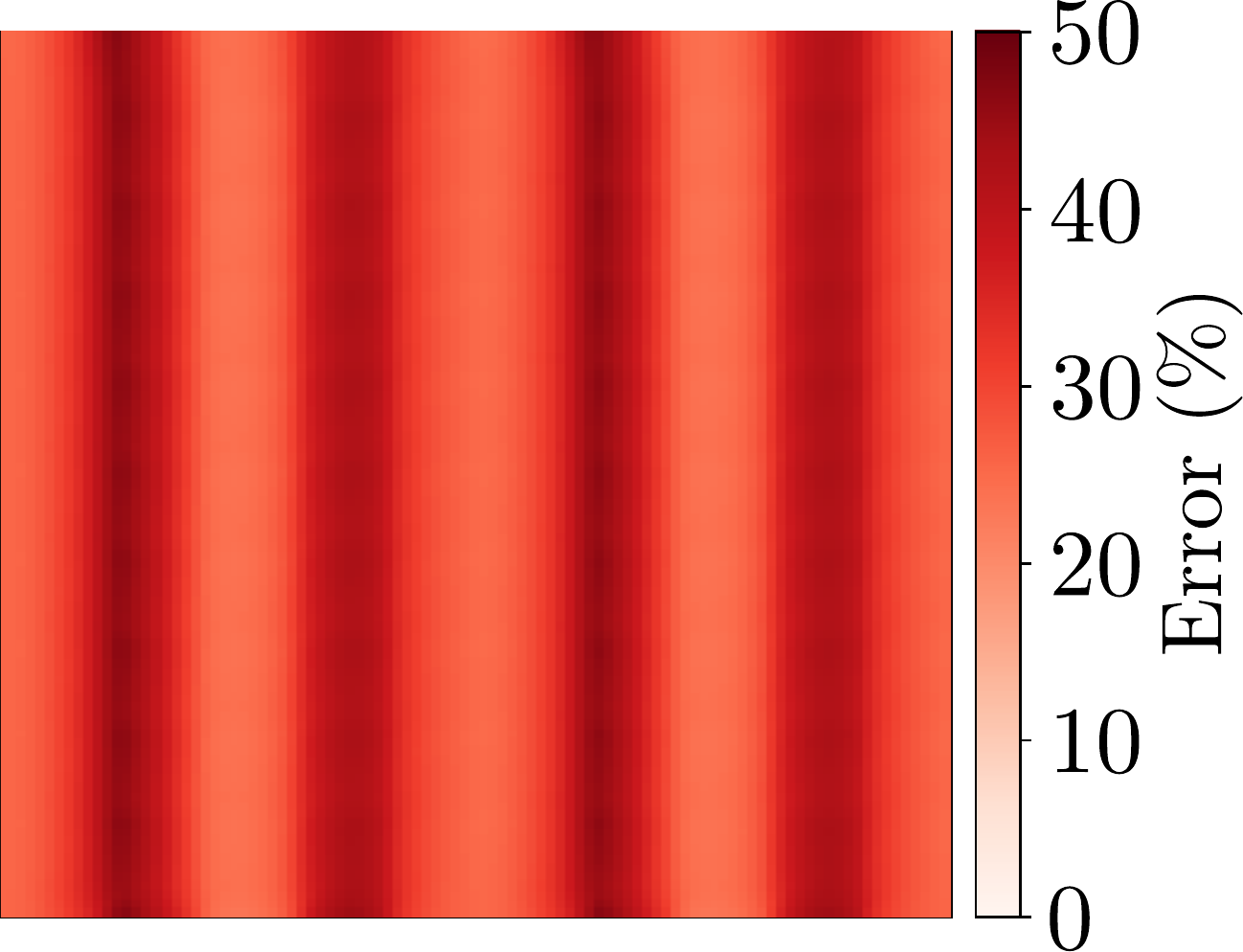}
      \label{sfig_pe_error}
    }
  \endgroup 
  \caption{Comparison of surface potential between linear and nonlinear models.
    Errors from the nonlinear PBE model are shown for the \protect\subref{sfig_le_error} LE and \protect\subref{sfig_pe_error} PE physical models respectively.
    Though both provide similar solutions, the result from the ALE model is a closer representation to the full PBE solution at the silicon surface.
    The offset used in \protect\subref{sfig_le_error} is obtained using the Boltzmann relation with dopant $\SI{10}{}^{19}$ \SI{}{\per\centi\meter\cubed}, silicon bulk, and a temperature of \SI{77}{\kelvin}.
  }
  \label{fig:model_errors}
\end{figure}

\Cref{fig:model_errors} compares the PBE and LE/PE models.
The LPBE model is neglected here due to its non-generality, only being representative for solutions centered around $u = 0$.
To compensate for the LE's neglect of charge density, an approximate offset was found by the Boltzmann relation and added to the LE solution, resulting in the augmented Laplace's Equation (ALE).

At the SiDB surface, both linear models converge to solutions of the same magnitude as the PBE model, with the ALE having smaller error.
For this reason, this work avoids invoking the time prohibitive nonlinear PBE, instead using the ALE to approximate the electric potential at the silicon surface.

\subsection{Verification with COMSOL}
\begin{figure}
  \centering
  \begingroup 
    \captionsetup[subfigure]{width=0.5\linewidth}
    \subfloat[PoisSolver]{%
      \includegraphics[width=0.45\linewidth]{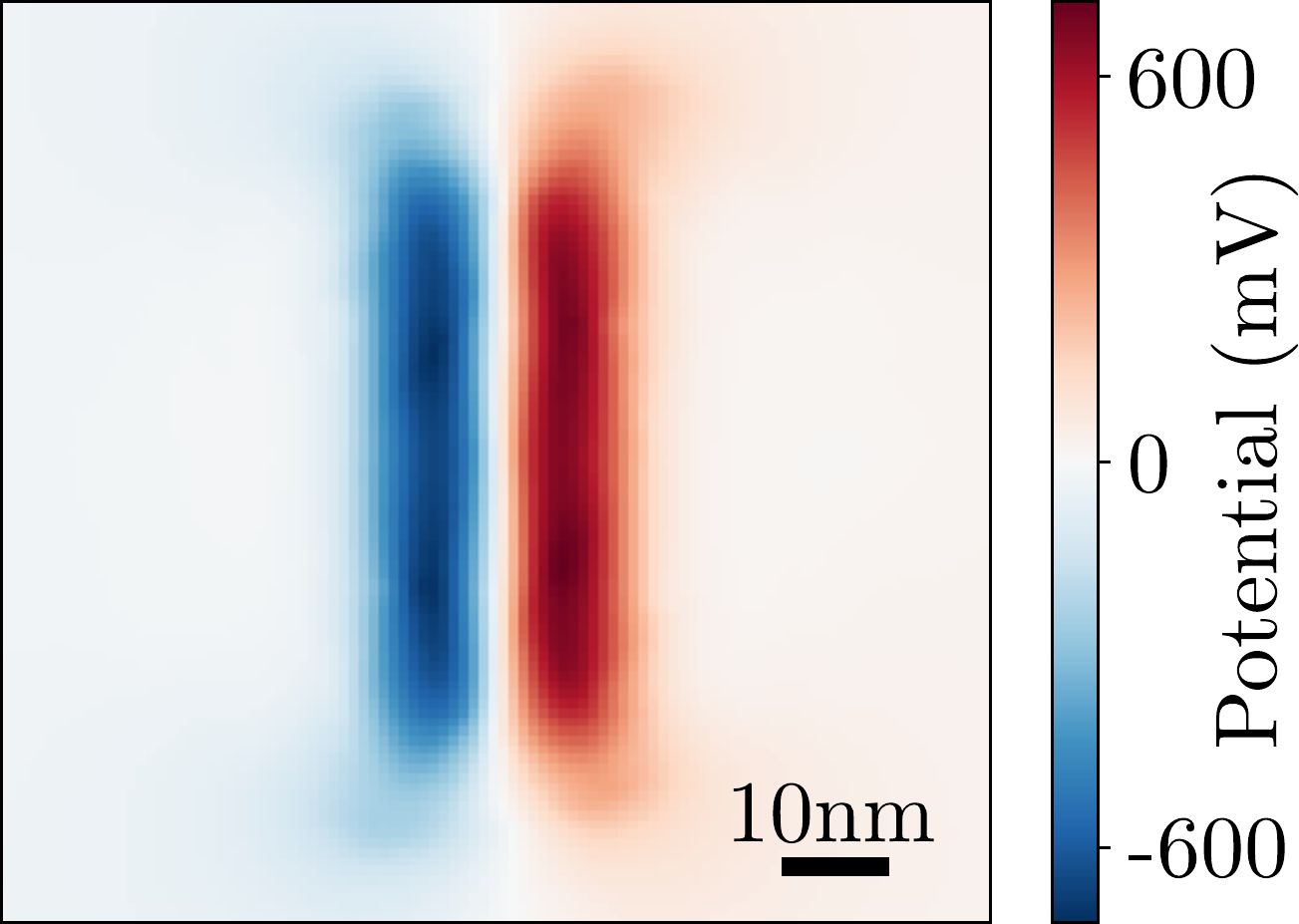}
      \label{sfig_poissolver}
    }
    \quad
    \subfloat[COMSOL]{%
      \includegraphics[width=0.45\linewidth]{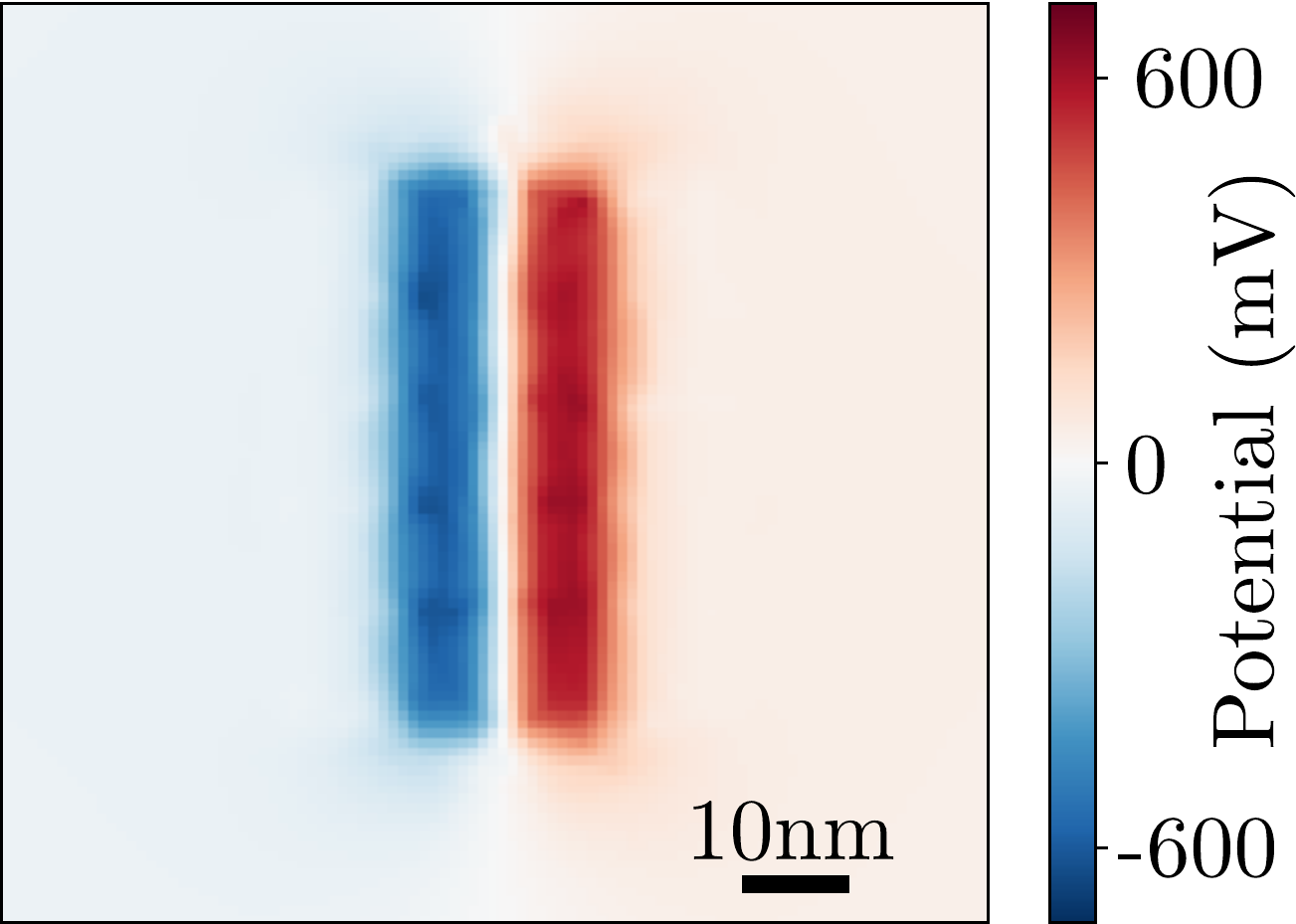}
      \label{sfig_comsol}
    }
    \\
    \subfloat[Difference]{%
      \includegraphics[width=0.425\linewidth]{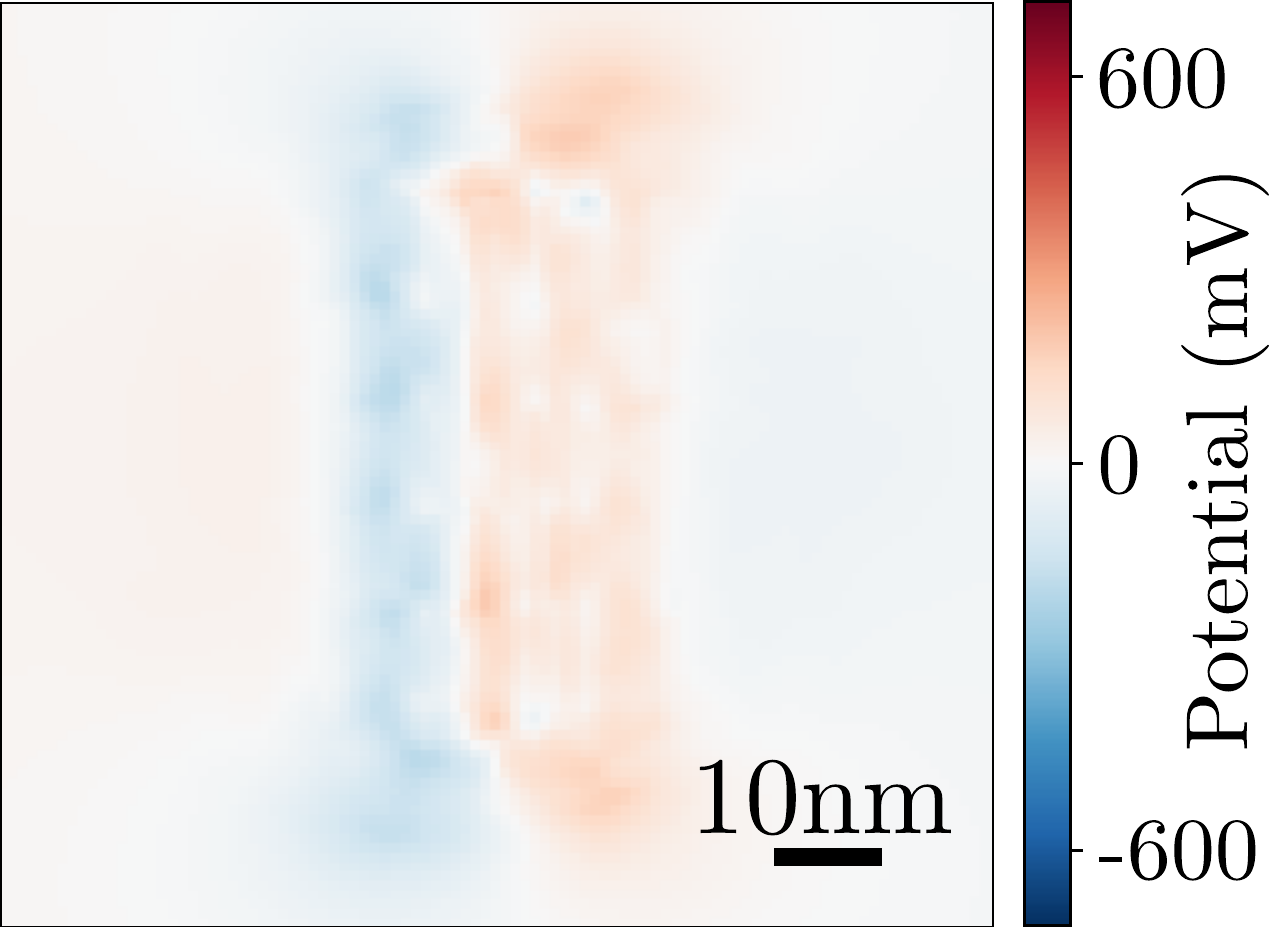}
      \label{sfig_diff}
    }
    \quad
    \subfloat[Schematic]{%
      \includegraphics[width=0.425\linewidth]{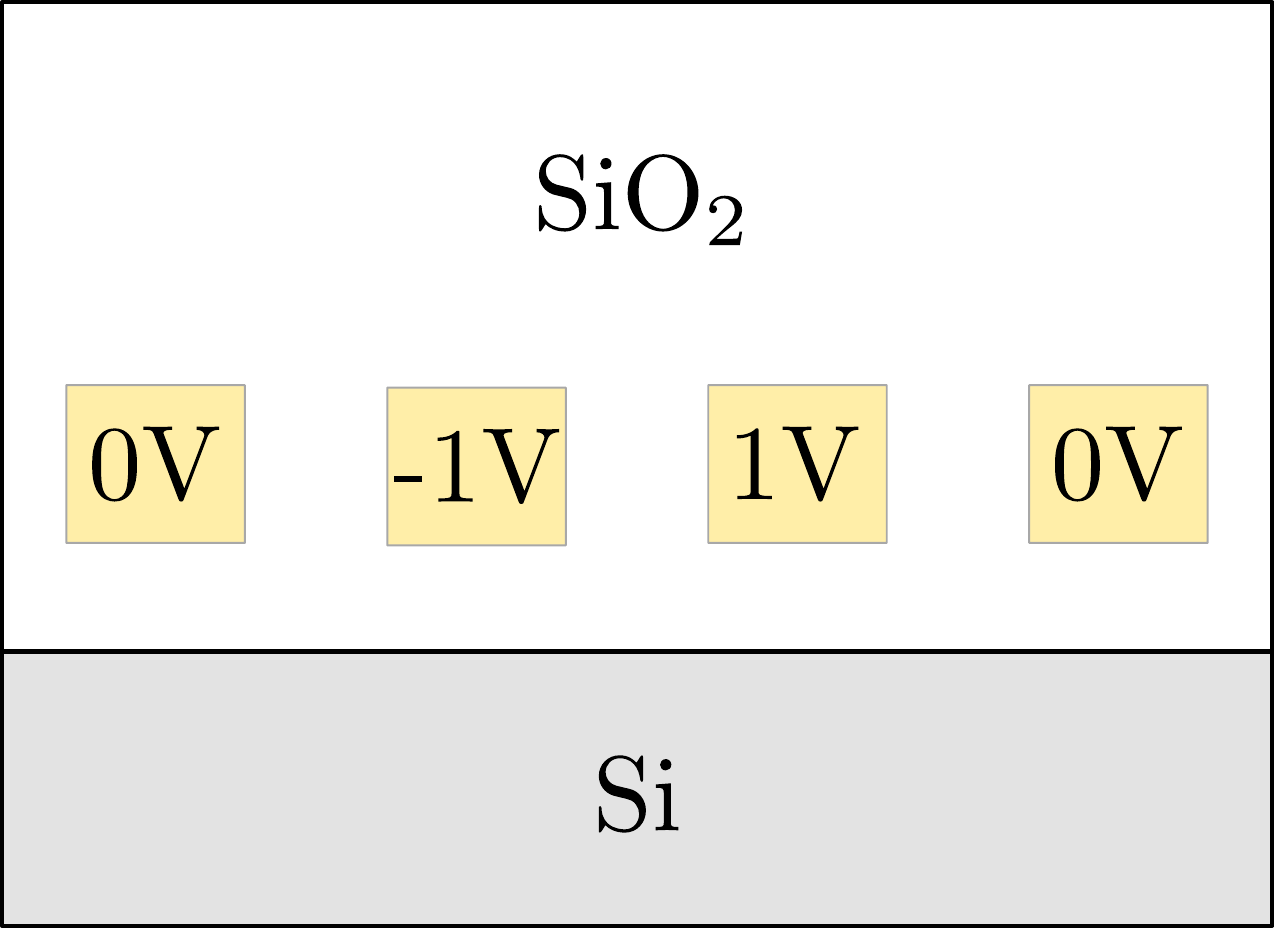}
      \label{sfig_schem}
    }
  \endgroup 
  \caption{Numerical comparison between PoisSolver and COMSOL results for a layout consisting of a row of electrodes, with two electrodes set at $\pm \SI{1}{\volt}$ surrounded by two grounded electrodes on either side, all suspended above a silicon substrate in dielectric.
  }
  \label{fig:comsol_compare}
\end{figure}

To verify that the results here are valid, we compare results for a four electrode layout from both PoisSolver and COMSOL, a commercial finite element general physics simulator.
\Cref{sfig_poissolver,sfig_comsol} show surface potentials for the system as found by each simulator.
\Cref{sfig_diff} shows the difference when comparing the two tools.
PoisSolver and COMSOL results are on the same order of magnitude, with a maximum $25\%$ error.

\subsection{Clocking network results}
Using PoisSolver, surface potential patterns arising from the clocking schemes in \Cref{fig:schemes_gui_layout} are simulated and visualised.
The electrode heights and clocking voltages are tuned such that a $\pm$\SI{100}{\milli\volt} potential swing appears on the surface, enough to trigger a SiDB state transition \cite{ng2020siqad}.
Example results for each clocking scheme are shown in \Cref{fig:schemes_gui_layout}.
Though \Cref{fig:schemes_gui_layout} shows only a single snapshot in time, full clock cycles can be simulated by repeatedly solving the system with different electrode boundary conditions.

\section{Circuit model of the clocking network} \label{sec_circuit}
\begin{figure}
  \centering
  \includegraphics[width=\linewidth]{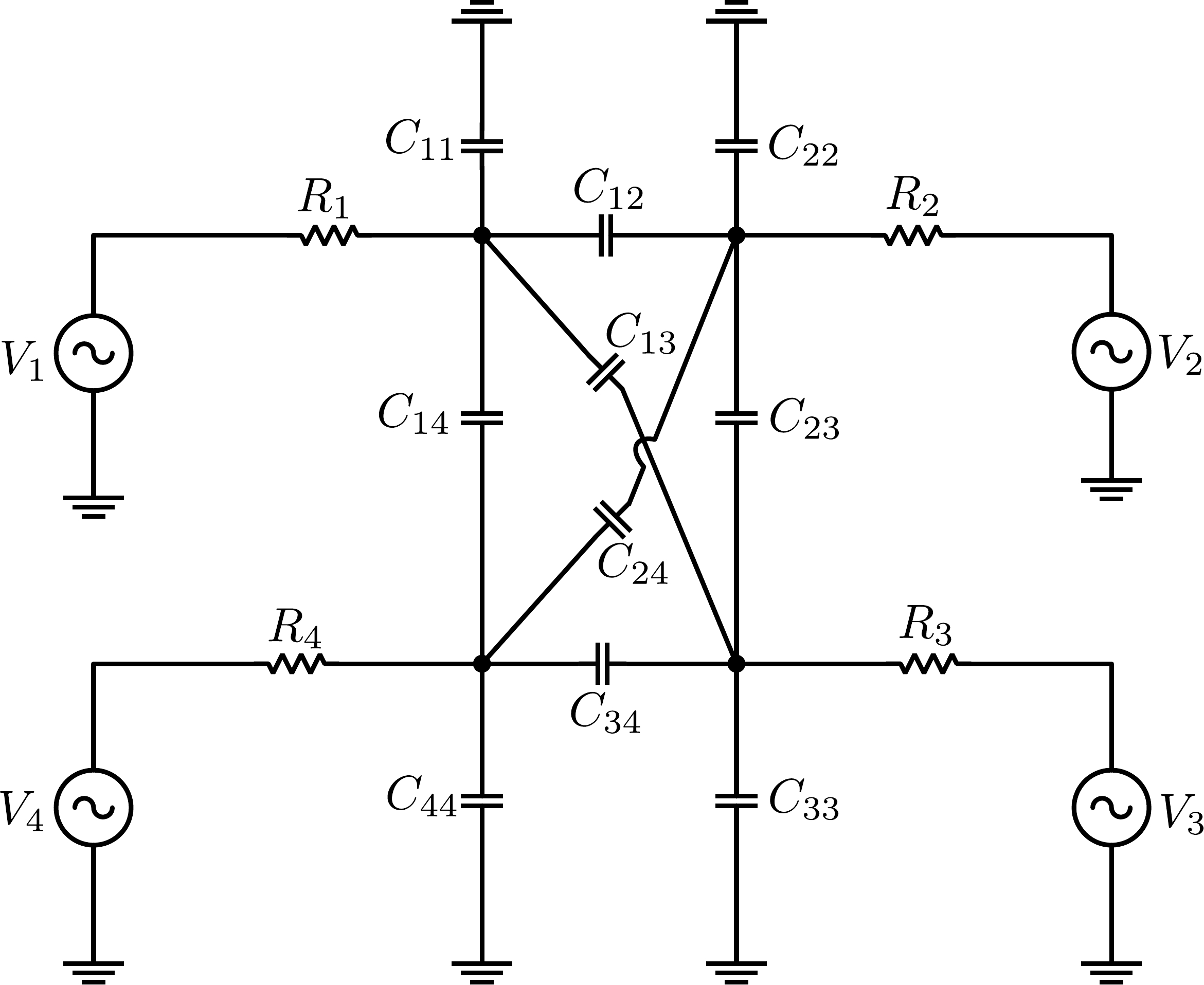}
  \caption{The circuit model for the four phase clocking network.
  The model uses $R_i$ to represent resistive losses, $C_{ij}$  for interelectrode capacitances, and $C_{ii}$ for self capacitances.
  }
  \label{fig:circuit_model}
\end{figure}

To model the power dissipation of the clocking network, the clocking system is mapped to a network of resistances and capacitances.
A schematic of the circuit model is shown in \Cref{fig:circuit_model}.
Each clocked electrode is modelled by a sinusoidal power supply with voltage amplitude $V_i$ and series resistance $R_i$.
Inter-electrode and self-capacitances are denoted by $C_{ij}$ and $C_{ii}$, respectively.
These capacitance and resistance values depend heavily on geometry and material properties.

\subsection{Capacitance estimation} \label{subsec_cap}
Many capacitance estimation techniques are geometrically constrained.
Rather than using an approximation for an assumed geometry, a field solver method \cite{lorenzo2011maxwell} is employed.
Briefly, this method solves for the 3D potential with PoisSolver to obtain sympathetic charges on electrodes.
These charges are then used to generate the Maxwell capacitance matrix, yielding approximations for all capacitances in \Cref{fig:circuit_model}.
Results from the field solver are within $1\%$ of parallel plate approximations.

\subsection{Resistance estimation} \label{subsec_res}
To approximate the effective resistance of an interconnected electrode network, the inverse Laplacian method \cite{ellens2011effective} is applied.
To start, the resistance of each straight resistor segment is estimated from the electrode geometry provided by SiQAD.
A graph is then created, and the Laplacian method is used to obtain the effective resistance between the highest point of the network, where signals are fed in, to the lowest point where clocking fields are generated.
The resistance from the inverse Laplacian method is within $2.1\%$ of a reference calculation.
This combined method reproduces results obtained by other approximations without any limitation on geometry.

\section{Power dissipation of four phase clocking schemes} \label{sec_power}
With the resistance and capacitance estimation methods from Section \ref{sec_circuit}, power estimates are obtained for the three clocking schemes.
Each scheme assumes the use of periodically tiled copper electrodes and SiO\textsubscript{2} dielectric with physical dimension constraints defined in \Cref{table_parameters}.

\begin{figure}
    \centering
      \includegraphics[width=\linewidth,height=6cm]{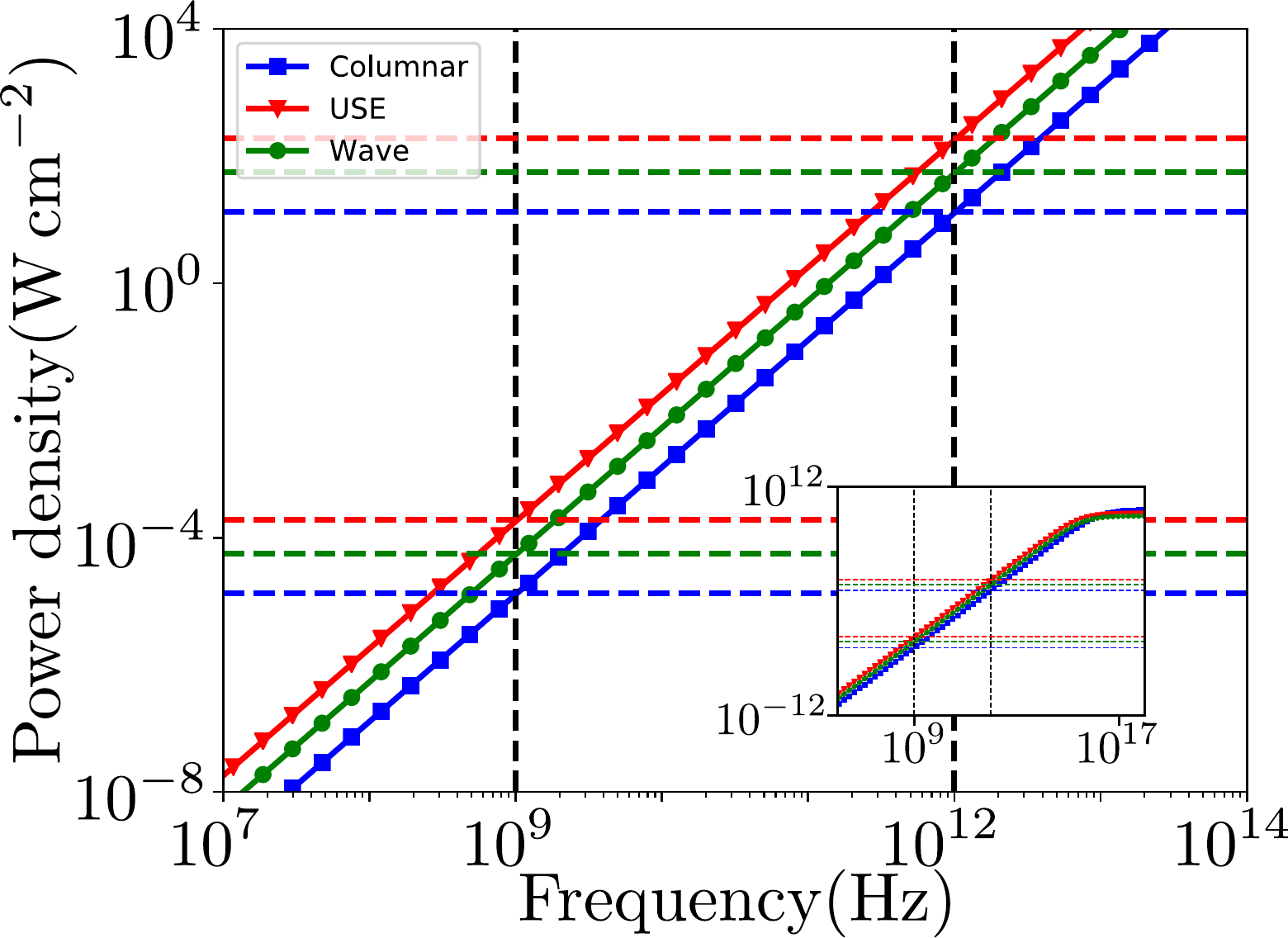}
      \caption{Frequency sweep of resistive power dissipation normalised to clocking area using copper electrodes with temperature \SI{77}{\kelvin}.
      Values shown are averages over all electrodes in the network.
      At \SI{1}{\giga\hertz}, the power draw of each clocking layout is estimated to be \SI{13.4}{}, \SI{56.3}{}, and \SI{189}{\micro\watt\per\centi\meter\squared} for the columnar, two dimensional wave, and USE schemes respectively.
      These powers increase to \SI{1}{}-\SI{10}{\watt\per\centi\meter\squared} range at \SI{1}{\tera\hertz}.
      }  \label{fig:power_plots}
\end{figure}

\Cref{fig:power_plots} shows frequency sweeps of parasitic power dissipation for each clocking scheme up to the \SI{100}{\tera\hertz} range.
The simulation accounts for temperature dependencies, using cryogenic resistivity data interpolated from \cite{white1959electrical}.
Temperature dependence of the dielectric is neglected here, being on the order of \SI{0.0001}{\per\kelvin} \cite{jie2012high}.
At the operating frequency of \SI{1}{\giga\hertz} and \SI{77}{\kelvin} (liquid nitrogen), the clock schemes dissipate power on the order of \SI{10}{}-\SI{100}{\micro\watt\per\centi\meter\squared}.
These power densities increase to \SI{1}{}-\SI{10}{\watt\per\centi\meter\squared} at \SI{1}{\tera\hertz}.

Limiting factors on the maximum frequency of these clocking schemes include dissipated heat, transmission line effects, and microstrip signal injection capabilities.
State-of-the-art cooling techniques can handle up to \SI{1}{\kilo\watt\per\centi\metre\squared} \cite{zhang2015water}, making these layouts viable up to the \SI{10}{\tera\hertz} range.
For high frequency signals, transmission line effects are non-negligible when conductor lengths are roughly $\frac{1}{10}$ the signal wavelength \cite{haytengineering}.
Under this rule of thumb, transmission line effects are expected to be negligible up to the \SI{100}{\tera\hertz} range.
\cite{zhu2016design} experimentally demonstrates the successful injection of \SI{0.1}{}-\SI{1}{\tera\hertz} signals on microstrip lines.
Under these constraints, the clocking schemes implemented here are expected to perform up to the \SI{1}{\tera\hertz} range.

The columnar network exhibits the lowest power dissipation, while the USE network has the highest power dissipation at any particular operating point, with the wave scheme falling somewhere in between.
These differences are attributed to each scheme's routing complexity, with complex schemes having higher interelectrode capacitances.

\begin{figure}
    \centering
    \includegraphics[width=\linewidth,height=6cm]{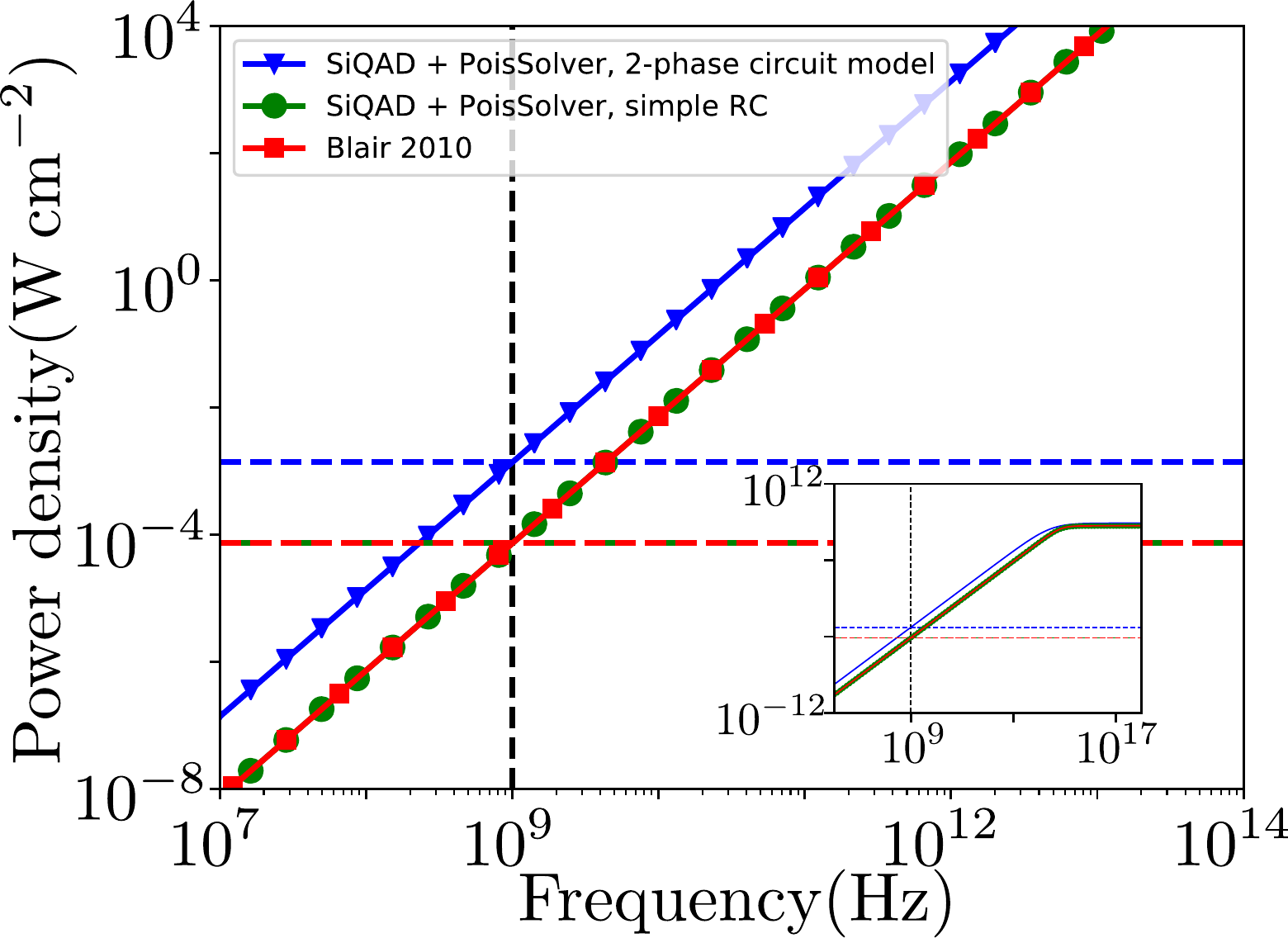}
    \caption{Numerical comparison between the results from simulation in SiQAD + PoisSolver and the approximations used in \cite{blair2010power} for the two phase columnar clocking layout.
    The length, width, distance, and pitch of the electrodes are \SI{1000}{\nano\metre}, \SI{50}{\nano\metre}, \SI{50}{\nano\metre}, \SI{100}{\nano\metre} respectively.
    The simulation is performed at \SI{293}{\kelvin} with copper electrodes.
    Using the two phase circuit model, the SiQAD + PoisSolver (blue) estimates power dissipation at \SI{1.40}{\milli\watt\per\centi\meter\squared} at \SI{1}{\giga\hertz}.
    With the simple RC model, estimated power dissipations are \SI{74.1}{} and \SI{74.0}{\micro\watt\per\centi\meter\squared} at \SI{1}{\giga\hertz} using SiQAD + PoisSolver (green) and from \cite{blair2010power} (red) respectively.
    }\label{fig:power_compare}
\end{figure}

\Cref{fig:power_compare} compares power figures to those in literature.
Under identical assumptions (adapting a simple two phase circuit model), PoisSolver reproduces the power results found in \cite{blair2010power}.
However, by switching to a more representative two phase model resembling \Cref{fig:circuit_model}, power estimates increase by a factor of 20.
As the interelectrode capacitance and resistance parameters are nearly identical in both models, the discrepancy is attributed to the inclusion of another path to ground through self capacitances.


\section{Comparison with device power consumption} \label{sec_comparison}
Although power densities investigated have been related to electrode clocking, the presence of SiDBs on the surface introduce other mechanisms of loss.
Figure 11 in \cite{timler2002power} relates the power consumption of a QCA cell to its kink energy.
Since QCA is a potential computing architecture that can be implemented in the SiDB platform \cite{wolkow2014silicon}, we use this as an estimate for power dissipated in the DB network itself. .
SiDBs incur relaxations of approximately \SI{200}{\milli\eV} \cite{northrup1989effective, rashidi2018initiating}, and an SiDB OR gate takes up an area of \SI{28}{\nano\meter\squared}.
With a cooling limit of \SI{1}{\kilo\watt\per\centi\metre\squared}, the trend of Figure 11 in \cite{timler2002power} suggests a maximum operating frequency of \SI{10}{\giga\hertz} for a SiDB system, two orders of magnitude slower than the limit of the clocking network.
These estimates suggest that the bandwidth of the clocking network will not bottleneck SiDB operating frequency.

With this model, the power dissipation of the electrode network is dominated by the power dissipation of SiDB devices.
However, the exact nature of SiDB power dissipation as electron populations and charge configurations change is not well understood.
\cite{todorovic2002diffusion} suggests that electrons along SiDB wires hop as polarons (a combination of a defect state and a trapped electron), possibly mitigating energy loss through lattice relaxation.

\section{Summary and conclusion}\label{sec_summary}
In this paper, PoisSolver was introduced as a 3D potential solver for the SiQAD environment.
Three clocking schemes from QCA were implemented in SiQAD and the resulting electrode networks were characterised with geometrically agnostic methods.
Using these characterisations, the power draw of four phase clocking schemes were estimated.
At \SI{77}{\kelvin}, the power draw of the columnar, wave, and USE layouts were found to be \SI{10}{}-\SI{100}{\micro\watt\per\centi\meter\squared} at \SI{1}{\giga\hertz} and \SI{1}-\SI{10}{\watt\per\centi\meter\squared} at \SI{1}{\tera\hertz}.
Although current estimations of SiDB device power overshadow clocking dissipations, these estimates may change as dissipation mechanisms for SiDBs become better understood.

\bibliographystyle{ieeetr}
\bibliography{paper.bib}

\begin{thebibliography}{10}

\bibitem{wolkow2014silicon}
R.~A. Wolkow, L.~Livadaru, J.~Pitters, M.~Taucer, P.~Piva, M.~Salomons,
  M.~Cloutier, and B.~V. Martins, ``Silicon atomic quantum dots enable
  beyond-cmos electronics,'' in {\em Field-Coupled Nanocomputing}, pp.~33--58,
  Springer, 2014.

\bibitem{huff2017atomic}
T.~R. Huff, H.~Labidi, M.~Rashidi, M.~Koleini, R.~Achal, M.~H. Salomons, and
  R.~A. Wolkow, ``Atomic white-out: Enabling atomic circuitry through
  mechanically induced bonding of single hydrogen atoms to a silicon surface,''
  {\em ACS nano}, vol.~11, no.~9, pp.~8636--8642, 2017.

\bibitem{anderson2014field}
N.~G. Anderson and S.~Bhanja, ``Field-coupled nanocomputing,'' {\em Lecture
  Notes in Computer Science (LNCS)}, vol.~8280, 2014.

\bibitem{huff2018binary}
T.~Huff, H.~Labidi, M.~Rashidi, L.~Livadaru, T.~Dienel, R.~Achal, W.~Vine,
  J.~Pitters, and R.~A. Wolkow, ``Binary atomic silicon logic,'' {\em Nature
  Electronics}, vol.~1, no.~12, p.~636, 2018.

\bibitem{ng2020siqad}
S.~S.~H. {Ng}, J.~{Retallick}, H.~N. {Chiu}, R.~{Lupoiu}, L.~{Livadaru},
  T.~{Huff}, M.~{Rashidi}, W.~{Vine}, T.~{Dienel}, R.~{Wolkow}, and K.~{Walus},
  ``Siqad: A design and simulation tool for atomic silicon quantum dot
  circuits,'' {\em IEEE Transactions on Nanotechnology}, pp.~1--1, 2020.

\bibitem{hennessy2001clocking}
K.~Hennessy and C.~S. Lent, ``Clocking of molecular quantum-dot cellular
  automata,'' {\em Journal of Vacuum Science \& Technology B: Microelectronics
  and Nanometer Structures Processing, Measurement, and Phenomena}, vol.~19,
  no.~5, pp.~1752--1755, 2001.

\bibitem{lent2003molecular}
C.~S. Lent, B.~Isaksen, and M.~Lieberman, ``Molecular quantum-dot cellular
  automata,'' {\em Journal of the American Chemical Society}, vol.~125, no.~4,
  pp.~1056--1063, 2003.

\bibitem{taucer2014single}
M.~Taucer, L.~Livadaru, P.~G. Piva, R.~Achal, H.~Labidi, J.~L. Pitters, and
  R.~A. Wolkow, ``Single-electron dynamics of an atomic silicon quantum dot on
  the h- si (100)-(2$\times$ 1) surface,'' {\em Physical review letters},
  vol.~112, no.~25, p.~256801, 2014.

\bibitem{rashidi2016time}
M.~Rashidi, J.~A. Burgess, M.~Taucer, R.~Achal, J.~L. Pitters, S.~Loth, and
  R.~A. Wolkow, ``Time-resolved single dopant charge dynamics in silicon,''
  {\em Nature communications}, vol.~7, no.~1, pp.~1--7, 2016.

\bibitem{pitters2011charge}
J.~L. Pitters, I.~A. Dogel, and R.~A. Wolkow, ``Charge control of surface
  dangling bonds using nanoscale schottky contacts,'' {\em ACS nano}, vol.~5,
  no.~3, pp.~1984--1989, 2011.

\bibitem{blair2018clock}
E.~Blair and C.~Lent, ``Clock topologies for molecular quantum-dot cellular
  automata,'' {\em Journal of Low Power Electronics and Applications}, vol.~8,
  no.~3, p.~31, 2018.

\bibitem{vankamamidi2008two}
V.~Vankamamidi, M.~Ottavi, and F.~Lombardi, ``Two-dimensional schemes for
  clocking/timing of qca circuits,'' {\em IEEE Transactions on Computer-Aided
  Design of Integrated Circuits and Systems}, vol.~27, no.~1, pp.~34--44, 2008.

\bibitem{campos2016use}
C.~A.~T. Campos, A.~L. Marciano, O.~P.~V. Neto, and F.~S. Torres, ``Use: A
  universal, scalable, and efficient clocking scheme for qca,'' {\em IEEE
  Transactions on computer-aided design of integrated circuits and systems},
  vol.~35, no.~3, pp.~513--517, 2016.

\bibitem{blair2010power}
E.~P. Blair, E.~Yost, and C.~S. Lent, ``Power dissipation in clocking wires for
  clocked molecular quantum-dot cellular automata,'' {\em Journal of
  computational electronics}, vol.~9, no.~1, pp.~49--55, 2010.

\bibitem{liu2012review}
W.~Liu, M.~O'Neill, and E.~E. Swartzlander, ``A review of qca adders and
  metrics,'' in {\em 2012 Conference Record of the Forty Sixth Asilomar
  Conference on Signals, Systems and Computers (ASILOMAR)}, pp.~747--751, IEEE,
  2012.

\bibitem{sicard2017introducing}
E.~Sicard, ``Introducing 14-nm finfet technology in microwind,'' 2017.

\bibitem{logg2010dolfin}
A.~Logg and G.~N. Wells, ``Dolfin: Automated finite element computing,'' {\em
  ACM Trans. Math. Softw.}, vol.~37, pp.~20:1--20:28, Apr. 2010.

\bibitem{alnaes2015fenics}
M.~S. Aln{\ae}s, J.~Blechta, J.~Hake, A.~Johansson, B.~Kehlet, A.~Logg,
  C.~Richardson, J.~Ring, M.~E. Rognes, and G.~N. Wells, ``The fenics project
  version 1.5,'' {\em Archive of Numerical Software}, vol.~3, no.~100,
  pp.~9--23, 2015.

\bibitem{geuzaine2009gmsh}
C.~Geuzaine and J.-F. Remacle, ``Gmsh: A 3-d finite element mesh generator with
  built-in pre-and post-processing facilities,'' {\em International journal for
  numerical methods in engineering}, vol.~79, no.~11, pp.~1309--1331, 2009.

\bibitem{lorenzo2011maxwell}
E.~D. Lorenzo, ``The maxwell capacitance matrix,'' {\em FastFieldSolvers},
  pp.~1--3, 2011.

\bibitem{ellens2011effective}
W.~Ellens, F.~Spieksma, P.~Van~Mieghem, A.~Jamakovic, and R.~Kooij, ``Effective
  graph resistance,'' {\em Linear algebra and its applications}, vol.~435,
  no.~10, pp.~2491--2506, 2011.

\bibitem{white1959electrical}
G.~K. White and S.~Woods, ``Electrical and thermal resistivity of the
  transition elements at low temperatures,'' {\em Philosophical Transactions of
  the Royal Society of London. Series A, Mathematical and Physical Sciences},
  vol.~251, no.~995, pp.~273--302, 1959.

\bibitem{jie2012high}
Y.~Jie, W.~Bo, H.~Zhi-Ling, L.~Ming-Ming, C.~Wen-Qiang, B.~Chuan, F.~Xiao-Yong,
  and C.~Mao-Sheng, ``High-temperature permittivity and data-mining of silicon
  dioxide at ghz band,'' {\em Chinese Physics Letters}, vol.~29, no.~2,
  p.~027701, 2012.

\bibitem{zhang2015water}
R.~Zhang, M.~Hodes, N.~Lower, and R.~Wilcoxon, ``Water-based microchannel and
  galinstan-based minichannel cooling beyond 1 kw/cm2$ $ heat flux,'' {\em IEEE
  Transactions on Components, Packaging and Manufacturing Technology}, vol.~5,
  no.~6, pp.~762--770, 2015.

\bibitem{haytengineering}
W.~H. HAYT and J.~A. Buck, ``Engineering electromagnetics. 2010,'' {\em
  INOSTROZA, F}.

\bibitem{zhu2016design}
H.-T. Zhu, Q.~Xue, J.-N. Hui, and S.~W. Pang, ``Design, fabrication, and
  measurement of the low-loss soi-based dielectric microstrip line and its
  components,'' {\em IEEE Transactions on Terahertz Science and Technology},
  vol.~6, no.~5, pp.~696--705, 2016.

\bibitem{timler2002power}
J.~Timler and C.~S. Lent, ``Power gain and dissipation in quantum-dot cellular
  automata,'' {\em journal of applied physics}, vol.~91, no.~2, pp.~823--831,
  2002.

\bibitem{northrup1989effective}
J.~E. Northrup, ``Effective correlation energy of a si dangling bond calculated
  with the local-spin-density approximation,'' {\em Physical Review B},
  vol.~40, no.~8, p.~5875, 1989.

\bibitem{rashidi2018initiating}
M.~Rashidi, W.~Vine, T.~Dienel, L.~Livadaru, J.~Retallick, T.~Huff, K.~Walus,
  and R.~A. Wolkow, ``Initiating and monitoring the evolution of single
  electrons within atom-defined structures,'' {\em Physical review letters},
  vol.~121, no.~16, p.~166801, 2018.

\bibitem{todorovic2002diffusion}
M.~Todorovic, A.~Fisher, and D.~Bowler, ``Diffusion of a polaron in dangling
  bond wires on si (001),'' {\em Journal of Physics: Condensed Matter},
  vol.~14, no.~49, p.~L749, 2002.

\end{thebibliography}

\end{document}